\documentclass[epj,nopacs]{svjour}
\usepackage{graphics}
\usepackage{amsmath}
\usepackage{amssymb}

\begin{document}
\hugehead

\title{Final results of magnetic monopole searches with the MACRO experiment}

\author{
{\rm The MACRO Collaboration} \\
M.~Ambrosio$^{12}$, 
R.~Antolini$^{7}$, 
G.~Auriemma$^{14,a}$, 
D.~Bakari$^{2,17}$, 
A.~Baldini$^{13}$, 
G.~C.~Barbarino$^{12}$, 
B.~C.~Barish$^{4}$, 
G.~Battistoni$^{6,b}$, 
Y.~Becherini$^{2}$,
R.~Bellotti$^{1}$
% ...
\\
}
\vspace{-5mm}
\institute{
\footnotesize{
1. Dipartimento di Fisica dell'Universit\`a  di Bari and INFN, 70126 Bari, Italy \\
2. Dipartimento di Fisica dell'Universit\`a  di Bologna and INFN, 40126 Bologna, Italy \\
3. Physics Department, Boston University, Boston, MA 02215, USA \\
4. California Institute of Technology, Pasadena, CA 91125, USA \\
5. Department of Physics, Drexel University, Philadelphia, PA 19104, USA \\
6. Laboratori Nazionali di Frascati dell'INFN, 00044 Frascati (Roma), Italy \\
7. Laboratori Nazionali del Gran Sasso dell'INFN, 67010 Assergi (L'Aquila), Italy \\
8. Depts. of Physics and of Astronomy, Indiana University, Bloomington, IN 47405, USA \\
9. Dipartimento di Fisica dell'Universit\`a  dell'Aquila and INFN, 67100 L'Aquila, Italy\\
10. Dipartimento di Fisica dell'Universit\`a  di Lecce and INFN, 73100 Lecce, Italy \\
11. Department of Physics, University of Michigan, Ann Arbor, MI 48109, USA \\
12. Dipartimento di Fisica dell'Universit\`a  di Napoli and INFN, 80125 Napoli, Italy \\
13. Dipartimento di Fisica dell'Universit\`a  di Pisa and INFN, 56010 Pisa, Italy \\
14. Dipartimento di Fisica dell'Universit\`a  di Roma "La Sapienza" and INFN, 00185 Roma, Italy \\
15. Physics Department, Texas A\&M University, College Station, TX 77843, USA \\
16. Dipartimento di Fisica Sperimentale dell'Universit\`a  di Torino and INFN, 10125 Torino, Italy \\
17. L.P.T.P, Faculty of Sciences, University Mohamed I, B.P. 524 Oujda, Morocco \\
18. Dipartimento di Fisica dell'Universit\`a  di Roma Tre and INFN Sezione Roma Tre, 00146 Roma, Italy \\
19. Department of Physics, Sant Longowal Institute of Engg. \& Tech., Longowal 148 106, India\\
$a$ Also Universit\`a  della Basilicata, 85100 Potenza, Italy \\
$b$ Also INFN Milano, 20133 Milano, Italy \\
$c$ Also IASF/CNR sez. Bologna, 40129 Bologna, Italy \\
$d$ Also Universit\`a  di Trieste and INFN, 34100 Trieste, Italy \\
$e$ Also Dipartimento di Energetica, Universit\`a  di Roma, 00185 Roma, Italy \\
$g$ Also Institute for Space Sciences, 76900 Bucharest, Romania \\
$h$ Macalester College, Dept. of Physics and Astr., St. Paul, MN 55105 \\
$i$ Also Department of Physics, Pennsylvania State University, University Park, PA 16801, USA \\
$l $Also Dipartimento di Fisica dell'Universit\`a  della Calabria, Rende (Cosenza), Italy \\
$n$ Also RPD, PINSTECH, P.O. Nilore, Islamabad, Pakistan \\
$o$ Also Dipartimento di Scienze e Tecnologie Avanzate, Universit\`a  del Piemonte Orientale, Alessandria, Italy \\
$p$ Also U. Minn. Duluth Physics Dept., Duluth, MN 55812 \\
$q$ Also Dept. of Physics, MIT, Cambridge, MA 02139 \\
$r$ Also Intervideo Inc., Torrance CA 90505 USA \\
$s$ Also Resonance Photonics, Markham, Ontario, Canada\\
}}
\date{Received: July 2002 }
\abstract{
We present the final results obtained by the MACRO
experiment in the search for GUT magnetic monopoles in the 
penetrating cosmic radiation, for the range  
$4\times 10^{-5}< \beta < 1$. Several
searches with all the MACRO sub-detectors (i.e. scintillation
counters, limited streamer tubes and nuclear track detectors) were 
performed, both in stand alone and combined ways. No
candidates were detected and a $90\%$ Confidence Level (C.L.) upper limit to the
local magnetic monopole flux was set at the level of
$1.4\times 10^{-16}$ cm$^{-2}$ s$^{-1}$ sr$^{-1}$. This result 
is the first experimental limit obtained in direct searches which
is well below the Parker bound in the whole $\beta$ range in which GUT
magnetic monopoles are expected.}


\maketitle

\section{Introduction}
\label{sec:intro}
Within the framework of Grand Unified Theories (GUT) of the
strong and electroweak interactions,
supermassive magnetic monopoles (MMs) 
with masses $m \sim 10^{17} \,$GeV/c$^{2}$ could have been
produced in the early Universe as intrinsically stable topological
defects at the phase transition in which a simple gauge
symmetry left
an unbroken U(1) group \cite{monorev}.
At our time they can be searched for in the penetrating cosmic radiation as
``fossil" remnants of that transition.
The detection of such a particle would be one of the most spectacular confirmation of GUT
predictions.

The velocity range in which GUT  
magnetic 
monopoles should be sought spreads over
several decades.
If sufficiently heavy ($m \gtrsim 10^{17} \,$GeV/c$^{2}$), GUT   magnetic monopoles
would be gravitationally bound to the galaxy with a velocity distribution
peaked at $\beta~=~v/c~\simeq 10^{-3}$ \cite{monorev}. 
MMs trapped around the Earth or the Sun are expected to travel with 
$\beta \simeq 10^{-5}$ and $\simeq 10^{-4}$, respectively.

Intermediate mass   MMs ($10^{15}< m < 10^{17}$) could have been produced in later
phase transitions in the early Universe
at a lower energy scale \cite{shafi};
lighter  magnetic monopoles, with masses around $10^{7}\div10^{15} \,$GeV/c$^{2}$,
would be accelerated to relativistic velocities in one or more coherent 
domains of the galactic magnetic
field, or in the intergalactic field, or in several astrophysical sites 
like a neutron star \cite{lightMM}.

Theory does not provide definite predictions on the   magnetic monopole abundance.
However, by requiring that MMs
do not short-circuit the galactic magnetic
field faster than the dynamo mechanism can regenerate it, a flux upper limit can
be obtained. This is the so-called Parker bound
($\sim 10^{-15} \,$cm$^{-2}$s$^{-1}$sr$^{-1}$ \cite{parker}), whose value sets
the scale of the detector exposure for  MMs search.
The original Parker limit was re-examined to take into account 
the almost chaotic nature of the galactic magnetic field, with domain lengths of
about $1\div10$ kpc; the limit become mass-dependent. An Extended Parker 
Bound (EPB) at the level of $1.2\times 10^{-16}(m/10^{17})$ cm$^{-2}$s$^{-1}$sr$^{-1}$
was obtained \cite{epb}.

MACRO was a large multipurpose underground detector located in the Hall B of the
Laboratori Nazionali del Gran Sasso (Italy); it was optimized for the search for GUT
magnetic monopoles with velocity $ \beta \geq 4\times 10^{-5} $ and with a
sensitivity well below the Parker bound.
The detector, which took data up to December 2000, is fully described
in \cite{primosm,techpap}.
It was arranged in a modular structure, it was divided into six
$12.6 \times 12 \times 9.3 \,$m$^3$ sections referred to as \textit{SuperModules} (SM),
each one with separate mechanical structure and electronics readout.
The detector's global dimensions were $76.5 \times 12 \times 9.3 \,$m$^3$ and its 
total acceptance for an isotropic flux of
particles was $\sim 10,000 \,$m$^2$sr.

\begin{figure}
\hspace{0.3cm}
\resizebox{0.43\textwidth}{!}{\includegraphics*{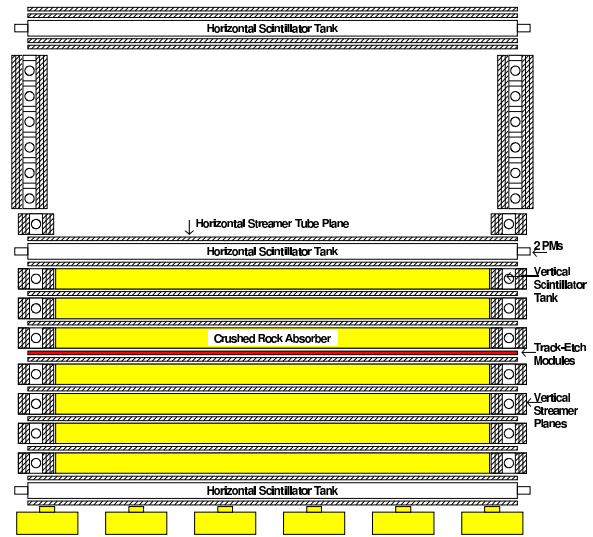}}
\caption{A cross-sectional view of the MACRO detector showing the layout of
         the 3 sub-detectors and of the rock absorber.}
\label{fig:intro1}
\end{figure}

Redundancy and complementarity were the primary features in designing the experiment.
Since we could not reasonably expect more than a few MMs during
the detector lifetime,
we deemed
crucial to have
multiple signatures and the ability to perform cross checks among various parts
of the apparatus. To accomplish this, the detector consisted of
three independent sub-detectors: liquid scintillation counters, limited streamer
tubes and nuclear track detectors, each of them with dedicated and independent hardware.
In Fig.~\ref{fig:intro1} a cross sectional view of the apparatus is shown.
Also visible are the 
seven horizontal 
absorber layers 
(which  set at $\sim 1 \,$GeV the
minimum energy threshold for throughgoing muons); 
notice the separation of the detector into a lower and an upper detector
(the \textit{attico}).

%
The background that magnetic 
monopole searches have to fight with is mainly due to
muons of the
cosmic radiation and natural radioactivity.
Thus large detectors  have to be installed in underground laboratories. 
At Gran Sasso
the minimum thickness of the rock overburden above the detector is
3150$\,$hg/cm$^2$. The cosmic radiation  muon flux in Hall B is 
$\sim 1 \,$m$^{-2}$h$^{-1}$,
almost a factor $10^6$ smaller than  that at the surface
(only muons with minimum energy of $\simeq 1.3 \,$TeV
can cross the mountain and reach MACRO) with an average muon residual energy
of $\sim 300 \,$GeV \cite{trd}.

The signatures of the passage of a GUT 
magnetic monopole across the
detector depend strongly on its velocity \cite{monorev,boeloss}.
For this reason, different hardware systems were designed and
operated to give optimum sensitivity  in different
$\beta$ values.
Different analysis strategies were also adopted, depending on the
$\beta$ range of interest and the subdetector(s)  used; the entire MM $\beta$ range 
($4 \times 10^{-5} < \beta < 1$) was thus covered.
In the different analyses we took into account both the
MM signatures and the background characteristics.

The unique property of a fast magnetic monopole 
with $\beta\geq 10^{-2}$ \cite{monorev,boeloss}
is its large ionization power compared either to the considerably slower 
magnetic monopoles or to the 
minimum ionizing electrically charged particles.
The searches for fast 
MMs look for large energy releases in the
detectors; the background  is mainly due to high energy muons 
with or without an accompanying electromagnetic shower.
On the other hand, slow 
magnetic monopoles should leave small signals spread over a
large time
window; a $\beta \sim 10^{-4}$ monopole could have a Time of Flight
(ToF) across the detector
as large as $\sim 1 \,$ms.
This implies the  use of specific
analysis procedures
that allow the rejection of the background mainly due to radioactivity
induced hits and possible electronic noise.

In this paper we report the final results of  several 
magnetic monopole
searches performed with MACRO (Sect.~\ref{sec:searches});
some early results were already published in \cite{monopap1}.
In each Section we discuss  the analysis criteria 
together with the results of each search; technical details may be found
in various papers fully describing the procedures and their application
to the first
data samples \cite{hong,phraserp,stmono,track,fastcomb}.
In Sect.~\ref{sec:global} the result of the combination of all the various
searches is reported.
In order to compare the MACRO 
results
to those of other experiments or to theoretical models,
we present upper limits
for an isotropic flux of bare MMs with
magnetic charge equal to one Dirac charge $g=g_D=e/2\alpha$,
(where $e$ is the electron charge and $\alpha$ the fine structure
constant) and nucleon decay catalysis cross section smaller than 1$\,$mb
\cite{monorev}. These aspects are discussed in Sect.~\ref{sec:discuss},  the conclusions  are given in Sect.~\ref{sec:conclu}.
A dedicate analysis for the search for magnetic monopoles accompanied 
by one or more nucleon decays along their path was also performed. 
The results are reported in a separate paper \cite{catalysis}.

\section{Experimental searches}
\label{sec:searches}
\subsection{Searches with Scintillators}
\label{intro}

The MACRO liquid scintillator system was organized in three layers of
horizontal and four layers of vertical counters, as shown in Fig.~\ref{fig:intro1}. 
For  atmospheric muons crossing the apparatus this system 
provided particle position,
energy deposition and ToF resolutions of about 11~cm,
1~MeV and 700~ps, respectively. 
The response of liquid
scintillators to heavily ionizing particles was studied both
experimentally \cite{ficenec} and theoretically \cite{Ahlen1,Ahlen2,ahlentarle} 
and their sensitivity to
particles with $\beta$ down to $\sim 10^{-4}$ was directly
measured.

\subsubsection{Wave Form Digitizer (WFD) analysis}
\label{wfd}
\newcommand{\almostless}{\hskip 4pt ^<_\sim\hskip 4pt}
Slow moving magnetic monopoles (in the range $10^{-4}\almostless
\beta \almostless$ $ 4.1 \times 10^{-3}$) were searched using dedicated
hardware, the Slow Monopole Trigger (SMT) and a 200 MHz
custom-made Wave Form Digitizer (WFD) system.
The main goal of the SMT was to remain sensitive over the entire
range of widths and amplitudes of pulses
(which could as well be just a train
of single photoelectron pulses lasting over several microseconds)
expected for slow moving magnetic monopoles while suppressing 
efficiently narrow (10 - 50~ns) pulses due to isolated radioactivity
and cosmic ray muons.
The SMT by itself offered only a hit register that recorded the
scintillator counters that satisfied the trigger conditions.
The photomultiplier tube (PMT) pulse shape information was recorded
by a custom made 200 MHz WFD system.
Redundant time-of-flight (TOF) information for every candidate event
was recorded by a stand-alone TDC.
The design and implementation of the triggering and recording
electronics used by this search has been described in detail
elsewhere \cite{primosm,techpap}.
\begin{figure}
\vspace{-0.3cm}
\resizebox{0.508\textwidth}{!}{\includegraphics*{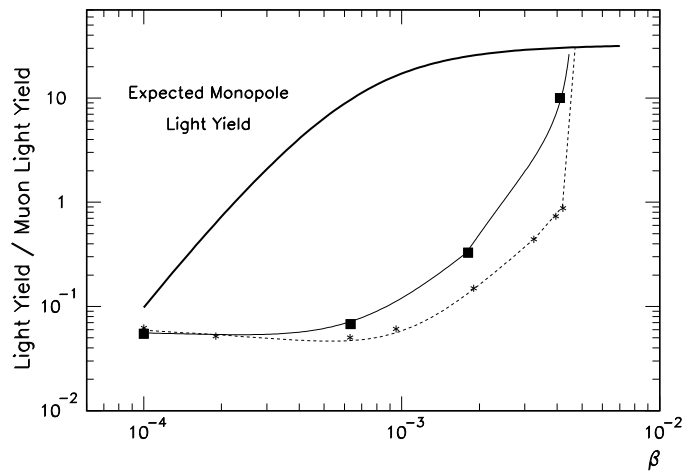}}
\caption{
      Results of the measurements of the Slow Monopole Trigger sensitivity 
	(black squares) and of the WFD analysis efficiency (stars) 
	plotted vs. $\beta$. The connecting lines serve only as a guide to the eye.	The expected light yield of a MM is also plotted.}
\label{smteff}       
\end{figure}

The heart of this slow MM search  was the analysis
of the digitized PMT signals as recorded by the 200MHz WFDs.
Several wave form analysis methods have been employed in
the search for slow MM searches in the initial
data taking and results from these have been
reported elsewhere \cite{hong}.
In this final search 
the SMT hardware \cite{primosm,techpap}
was used as the pattern recognition method for identifying a magnetic
monopole candidate event  \cite{sophia}.

The sensitivity of the trigger hardware as well as of the entire slow
MM analysis was measured experimentally in situ
and multiple times throughout MACRO's live time via LED light injection
in each of the scintillation counters.
The LED calibration system \cite{primosm,techpap}
allowed the generation of a magnetic monopole-like signal
in the detector of arbitrary direction, ionization
yield and velocity.
A grid of LED pulse widths and heights allowed the
generation of MM-like PMT pulses on a channel-by-channel
basis. The corresponding $\beta$-range
is $10^{-4} < \beta < 5\times 10^{-3}$ and the light
yield from few photoelectrons up to several times
the yield of a muon.
The generated wave forms were recorded by the WFD
system independently of whether the SMT fired or not.
Then, an off-line wave form analysis (the same used to analyze
the real events) established the ratio of the events that the SMT 
had selected over the ones expected.
This yielded the trigger sensitivity as a function of the particle's
velocity and it is shown in Fig.~\ref{smteff}.
More than $95\%$ of the detector's channels exhibited efficiency
above $99\%$ for light yields greater or equal the ones defined
by the curve.
The generated MM-like signals were then fed into the analysis wave form selection algorithm in discussion.
The stars in Fig.~\ref{smteff} show the analysis efficiency, when
more that $99\%$ of the simulated MM-like wave forms
--corresponding to lights yields and velocities across the
expected range--
were successfully identified by the analysis algorithm.

There were three main instrumental sources which cau\-sed the
hardware trigger to be less efficient than a software
one: (a) electronics were operating at a very low
threshold (2~mV) making them susceptible to ground loops
and low amplitude interference; (b) electronics suffered
from intrinsic pulse streching that may overestimated
pulse duration; (c) helium and hydrogen contamination in the PMT
envelope increased its activity,
leading sometime to overestimate of pulse duration.
A software pattern recognition method corrects for them,
and the efficiency of this algorithm is folded in the trigger
efficiency already discussed. 

Magnetic monopole candidate events were selected as follows.
The hardware trigger (SMT) was required to be present in at least two of the
detector faces and within 1msec.
This rejects mainly noise and radioactivity induced events.
The two layers requirement reduces by $17\%$ the
detector's acceptance.
The TOF requirement of 1~msec 
was dictated by the hardware, namely
the mode of operation of the WFD system \cite{primosm,techpap}.
The presence of the scintillator's fast particle trigger
was then used as a veto.
The fast coincidence (1~$\mu$s) between faces required by this trigger
selects particle tracks corresponding to $\beta$
well above the SMT's sensitivity (Fig.~\ref{smteff}).
It rejects mainly cosmic ray muons that triggered 
the SMT either due to pulse stretching or afterpulsing.
A final loose cut on the hits per face (less than 4) and number of layers
(less than 5) was applied
to reject events induced by electronic
noise, multi-muon or muon shower events.
None of these requirements is expected to affect the sensitivity
to slow MMs while any effect on the acceptance has
been taken into account in the Monte Carlo.
After applying the above cuts on the dataset of $\sim$4.75 years
of data-taking and $\sim$28 million SMT's, there were 35901 events left.
These were attributed to cosmic ray muons ($\approx 1/3$) 
and electronic noise ($\approx 2/3$) leaking out of our 
fast particle and loose noise/high multiplicity veto.
These events were further analyzed by our wave form analysis
method.

The wave form analysis required that PMT signals satisfied the
pattern recognition criteria on an end-to-end, face-to-face basis. 40 events survived. (There was an initial running of the detector
that suffered from wave form memory saturation and effectively
wave form loss. 14 events involving primarily large ionizations
were allowed to pass this stage of the analysis).

The 40 remaining candidates were 
individually analyzed using all available information from the
scintillator sub-system (the wave form shape, timing information from the wave forms, 
timing information from the precise muon system -ERP- and from the SMT 
TOF system) \cite{techpap}.
None of the events was compatible with the passage of a slow MM.
Instead, based on information available from both the 
scintillator and the streamer tube sub-systems, the 40 events
were classified as follows:

\begin{itemize}
\item{25 as bipolar electronic noise;
      using primarily the WFD information, they are distinguished from a 
      genuine MM signal due to the preciable amount 
      of positive content in the waveforms (PMTs operated
      with negative voltage).} 
\item{10 as muon-induced showers;
      using primarily the fast muon and occasionally WFD TOF
      information, these events reflect fast crossing of the
      apparatus that failed to fire the fast particle veto.}
\item{4 as probably spurious electrostatic discharges of the LED system;
      using primarily geometrical arguments reflecting the inconsistency
      of the event with the passage of a monopole crossing the detector at a 
      straight line at a constant velocity, and finally,}
\item{the ``spokesmen even''.
	This event belongs to a particular
	run in which a LED-induced event (simulating a close to a realistic 
	$\beta \sim 10^{-3}$ MM) was intentionally and secretly generated
	to test the efficiency of our analyses to detect a MM signal.
	The two analyses that should have been sensitive to the faked 
	``spokesmen'' MM (the present analysis, and the PHRASE one),
	found the event. 
	The ``spokesmen event'' was a LED-generated event and was subject
	to the limitations the LED system had in generating a MM-like pulse, namely its slow rise-time.}
\end{itemize}
No event was compatible with the passage of a single slow magnetic 
monopole.
The analyzed data were collected in the period July 1995 -- May
2000, corresponding to $\sim 4.3$ live-years.  The detector performance 
during the data-taking period was monitored on a run-by-run basis. 
The total exposure was
$\int A dt = 0.9 \times 10^{16}~{\rm cm}^{2} \, {\rm s} \, {\rm sr}$.
By taking into account the trigger and
analysis efficiencies we obtained a $90 \, \%$ C.L. flux upper limit of
$2.5\times 10^{-16}$~cm$^{-2}$s$^{-1}$sr$^{-1}$ 
for $10^{-4} < \beta < 4.1 \times 10^{-3}$
(curve ``WFD" in Fig.~\ref{fig:individuali}) \cite{sophia}.

\subsubsection{The PHRASE analysis}
\label{phrase}

\begin{figure}
\vspace{-0.3cm}
\resizebox{0.51\textwidth}{!}{%
\includegraphics*{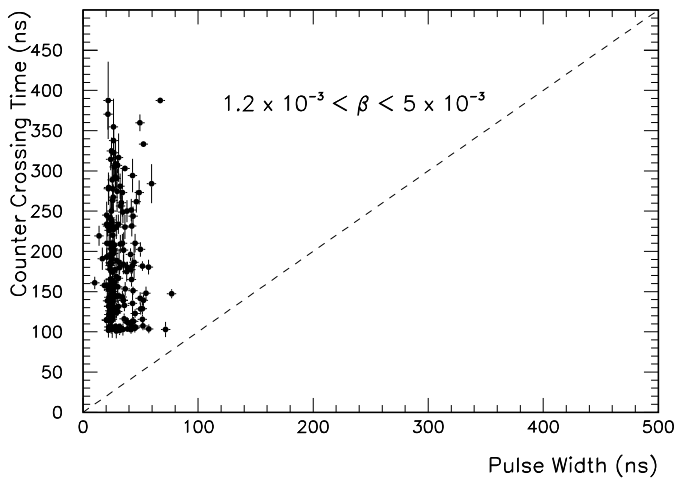}}
\caption{Calculated
         crossing time in a scintillation
         counter versus pulse width for PHRASE monopole
         candidates in the range $1.2 \times 10^{-3} < \beta < 5 \times 10^{-3}$
         (for clarity reasons only 40\% of the candidates is shown).
         The diagonal line
         indicates the minimum pulse width expected (corresponding to
         a MM with $\beta \simeq 5\times 10^{-3}$)
         for $15~{\rm cm}$ path length. No candidates meet or exceed this expectation.}
\label{figslow}
\end{figure}
Magnetic monopoles in the velocity range 
$1.2 \times 10^{-3} < \beta < 10^{-1}$ were
searched for using the   PHRASE
(Pulse Height Recorder And Synchronous Encoder)  system. 
It was a trigger and energy reconstruction processor which
generated a trigger condition when the energy deposition in a scintillator
was larger than a preset threshold, $E_{th} \sim 7~{\rm MeV}$ \cite{primosm,collasso,sorgente}.

The energy deposition was reconstructed by integrating
the photomultiplier pulses (recorded by a $100~{\rm MHz}$ Wave Form
digitizer system), taking into account their relative amplitudes, the
response functions of the scintillation counters and the liquid scintillator,
photomultiplier and electronic saturation effects. All these points
were studied in detail by using laser light of various intensities
and  atmospheric muon pulses; appropriate algorithms were developed for
correcting any nonlinearity which  may be present in large
energy loss rates as those expected from magnetic monopoles  in the analyzed $\beta$-range.
The analysis technique is described in detail in \cite{phraserp}. 

The  magnetic monopole candidate selection 
required hits in
two scintillator layers and a minimum energy deposit of $10~{\rm MeV}$ in
each of them, spread out over no more than four adjacent boxes (this cut
removed most of the natural radioactivity background but not atmospheric muons). 
The $100~{\rm MHz}$
sampling frequency of the PHRASE Wave Form digitizers and the $10~{\rm MeV}$
energy cut defined the minimum MM velocity which this search was
sensitive to: $\beta_{min} = 1.2 \times 10^{-3}$. The velocity of a particle
was reconstructed by using the TOF between the crossed scintillator layers;
a minimum distance of $2~{\rm m}$ between the hits was required to ensure
accurate  timing reconstructions. The timing and position uncertainties produced a
tail in the   muon velocity distribution which  is peaked at $\beta = 1$. 
Atmospheric muons were rejected by setting
the $\beta$ upper limit for the  magnetic monopole analysis to $0.1$. Finally,
a minimum path length of $15~{\rm cm}$ was required in each scintillation
counter (the typical path length for a crossing cosmic ray muon was
$> 20~{\rm cm}$). All geometrical cuts were taking into account
in computing the
detector acceptance.

The candidates surviving the selection (a few thousands in $\sim 11$ years of
running) were grouped in two $\beta$ overlapping sub-ranges, 
medium ($1.2 \times 10^{-3} < \beta < 5 \times 10^{-3}$) 
and high ($4 \times 10^{-3} < \beta < 10^{-1}$)
velocities. 
This was motivated by the fact that the MM light yield and
crossing time (with respect to muons) differ significantly
in these two regimes.
Crossing muons released $\sim 40~{\rm MeV}$ in each traversed
scintillation  layer, with a pulse width of $\approx 35~{\rm ns}$ 
due to the convolution between the time profile of the scintillation
light emission and propagation and the PMT time jitter.
The pulse width expected from a magnetic monopole in the medium $\beta$ range,  
corresponding to
the time needed to cross a counter, is $100 \div  400~{\rm ns}$. 
In the high $\beta$ range the expected
counter crossing time  is $6 \div 120~{\rm ns}$; thus the corresponding
pulse width  is comparable to that of a muon. The
expected energy loss rate, for MMs in the medium $\beta$ range,  is $10 \div 30$ times 
larger than that of a muon, 
and $30 \div 60$ times for  MMs in the higher
$\beta$ range.

For candidates with $1.2 \times 10^{-3} < \beta < 5 \times 10^{-3}$ we
compared the  measured pulse width with the expected counter crossing time.
The counter crossing time was calculated from the measured $\beta$, 
assuming a minimum pathlength in the scintillation counter of 15~cm 
(Fig.~\ref{figslow}).

For a 
particle crossing the detector with the measured velocity these two numbers
must be consistent. 
No final candidates
satisfied this condition  with the exception of the LED simulated ``spokesmen"
magnetic monopole introduced in the data (see sec.~2.1.1).

\begin{figure}
\vspace{-0.3cm}
\resizebox{0.51\textwidth}{!}{%
\includegraphics*{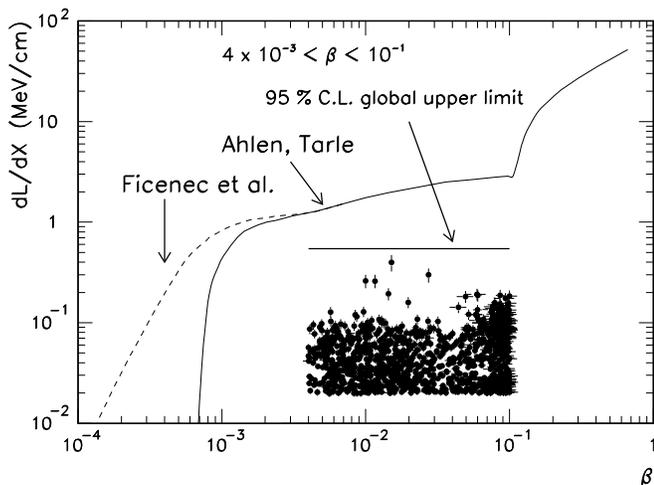}}
\vspace{-0.7cm}
\caption{  Energy loss in the scintillator versus velocity for PHRASE events in the range
         $4 \times 10^{-3} < \beta < 10^{-1}$ (for clarity
         only 40\% of the candidates
         is shown). The calculated MM energy loss
         rates from \cite{ficenec,ahlentarle} are also given, 
         as a $95 \, \%$
         C.L. upper limit on the maximum energy loss  of the candidates.}
        
\label{figfast}
\end{figure}

For candidates with $4 \times 10^{-3} < \beta < 10^{-1}$ we
compared the energy loss rate (computed with the
assumption of a fixed $15~{\rm cm}$ path length in the scintillator)
with that  expected for a MM
\cite{ficenec,ahlentarle} 
(see Fig.~\ref{figfast}).  
%
%
The $\beta$ range was extended down to $4 \times 10^{-3}$ to obtain an
independent  cross-check of the rejection based on the counter crossing
time criterion for the candidates at the boundary between the two
sub-intervals. 
In no case the energy deposition of a candidate was
consistent with that expected for a monopole of the 
same apparent velocity.

In this search magnetic monopole candidates were categorized 
on the basis of the streamer tube and
scintillator information. No event exhibits the pattern of a
single particle crossing the detector at the measured velocity;
the velocity values were artifacts due to uncommon classes
of events. A fraction of candidates ($\sim 10 \,\%$) comes from
occasional timing errors, which produced an apparent low value
for the velocity of a cosmic ray muon crossing the apparatus. In a larger
fraction of cases ($\sim 20 \%$), the low velocity value comes from
accidental coincidences, within the allowed time window, between
radioactivity  hits and/or  a cosmic ray muon occurring in different
scintillator layers.  The remaining candidates are due to cosmic ray
muons which crossed a scintillator layer and stopped just before
reaching a second one. The decay electron produces a hit in this second
layer, with a typical $\sim 2~\mu {\rm s}$ time delay (corresponding to
the muon life time) with respect to the first hit.

This analysis was applied to the whole data set collected 
from October 1989 to June 1995 with the lower part of the detector, and 
with the full detector   up to December 2000.
The detector acceptance was computed by a Monte Carlo simulation, 
which takes into account the fraction of apparatus effectively in 
acquisition in each individual run. 
The total exposure was
$\int A dt = 1.2 \times 10^{16}~{\rm cm}^{2} \, {\rm s} \, {\rm sr}$.
By taking into account the trigger and
analysis efficiencies we obtain a $90 \, \%$ C.L. flux upper limit
$\Phi \le 2.2 \times 10^{-16}~{\rm cm}^{-2} \, {\rm s}^{-1} \, {\rm sr}^{-1}$ 
for $1.2\times10^{-3}<\beta<10^{-1}$, which is
represented by curve ``PHRASE" in Fig.~\ref{fig:individuali}.

\subsection{Searches with Streamer Tubes}
\label{StreamerH}

The streamer tube system was designed to be effective in the search
for magnetic monopoles in a wide velocity range, $10^{-4} < \beta < 1$.
For this purpose we used
a gas mixture containing helium and
n-pentane.
Helium is necessary to exploit the Drell effect \cite{drellpate}: the passage
of a magnetic monopole with $1.1\times 10^{-4}< \beta < 10^{-3}$ through a 
helium atom leaves it in a excited metastable state.
The n-pentane can then be ionized by collision with excited helium atoms
(Penning effect). The high cross section for the processes ensures 100\%
efficiency in detecting bare MMs with this gas mixture.

In the higher velocity region ($\beta > 10^{-3}$), where the assumptions 
used in the Drell-Penning effects do not apply, the standard ionization mechanism
expected for a MM ensures an energy release several orders of magnitude
higher than that due to minimum ionizing particles.
The charge collected on a wire has a logarithmic
dependence on the energy released inside the active cell \cite{gbation,laser},
so that a charge measurement allowed one to distinguish 
between MMs and muons.

Horizontal  streamer tubes were equipped with readout strips in order to provide a three
dimensional event reconstruction. The memory depth of the readout electronics 
was large enough to store signals for monopoles with $\beta$ down to $10^{-4}$.
In order to retrieve spatial coordinates, signals were shaped at $\sim 550 \ \mu$s
and sent to parallel-in/serial-out shift register chains for readout. Analog ORs
of wire signals (covering 1 m for central planes and 0.5 m for lateral planes) were fed to 
the Charge and Time Processor (QTP) system for arrival time and charge measurement 
of the streamer signals \cite{primosm,techpap}.
The trigger was based on the assumption that a heavy MM crossed the 
apparatus without any appreciable change in its direction and speed, due to its large
kinetic energy. In designing the trigger
logic, particular attention was  given to avoiding the possibility 
that relativistic decay products from magnetic monopole induced nucleon decay, 
could cause dead-time 
in the apparatus before the  MM generated the trigger \cite{ivan}.

Horizontal and vertical  streamer tube planes were handled by two 
independent trigger systems. 
Monte Carlo simulations have shown that they were basically 
complementary. 
Two analysis were performed using data collected with the
two different triggers; all streamer tubes were used 
for the event reconstruction.
Both analyses were based on the search for single tracks
and on the measurement of the particle velocity by using the time information
provided by the  streamer tubes (maximum time jitter $\sim 600\,$ns) \cite{stmono}. 
Detailed investigations were performed in order to check that both trigger logics
and analysis efficiencies were independent of the particle velocity.

\subsubsection{Search with horizontal streamer tubes}
This analysis technique is described in details in \cite{stmono}.
The assumpition made in the trigger design allows 
the trigger logic to be sensitive to any massive 
particle able to produce a signal in the streamer tubes.
The streamer tubes time resolution allows identification of
slow particles by measuring their time of flight across the apparatus (``time track").
However, because of time jitter and afterpulsing in the streamer tubes, 
particles faster than $5\times10^{-3} c$ could be confused with cosmic ray muons.

The analysis was based on the search for 
single space tracks in both wire and strip views and on the measurement of the velocity of
the candidates.
The width of the temporal
window is so large that on average three spurious hits 
(due to the 40 Hz/m$^{2}$ background on streamer tubes) are
present in a $12 \times 12$~m$^{2}$ plane per event.
The trigger logic selected events with an alignment in $z$ (the vertical
coordinate) versus time $t$ of at least
seven streamer tube planes in one or more of 320 time windows (called $\beta$-slices).
Each $\beta$-slice covered a TOF window of 3 $\mu$s and provided a rough 
estimate of the particle time of flight across the apparatus.

The TOF  value provided by the trigger allowed a discrimination of  the streamer tube 
hits from a real track, from the spurious hits recorded in the $\sim 500$ $\mu$s memory depth
of the readout electronics.
If a space track was found, a more refined alignment in the $z-t$ temporal view
was required (``time track").
This rejected effectively the  background from radioactivity and the background 
from relativistic
muons combined accidentally with some radioactivity hits. 

The muon provided the spatial track, while the radioactivity hits may confuse 
the time tracking algorithm. Such events 
were rejected using a cut on the measured velocity.
In fact, they were tagged with both the first $\beta$-slice (that of relativistic muons)
and with a higher $\beta$-slice value (that of slow particles).
For particles crossing
the lower detector (path length $\sim$ 5~m) the above condition corresponds to a cut for all 
particles with  
$\beta >\beta_{cut}=$ (5~m/3~$\mu$s)/c = 5$\times 10^{-3}$.

The sensitivity of the trigger to relativistic muons provided a simple and efficient way to
estimate the overall efficiency of this analysis. Muons crossing at least seven
horizontal streamer tube planes survived all the analysis steps and were only rejected by 
the final $\beta$-cut. 
The efficiency $\varepsilon _a$  of the analysis was then estimated by the 
ratio of the  number of 
reconstructed muons to the expected  number of single muons crossing at 
least seven planes.
A further possible source of inefficiency  were failures in the trigger 
circuits $\varepsilon _t$.  This effect was not accounted for by 
the above procedure,
because muons were concentrated in 
the  first $\beta$-slice. 
The efficiencies $\varepsilon _a$ and $\varepsilon _t$, were completely described in \cite{stmono}. 
%

This analysis used the data sample collected with the horizontal streamer tube trigger
from January 1992 to September 2000, integrating a livetime of 8.1 years. 
The overall  average efficiency was $74 \% $. 
The detector acceptance, computed by a Monte Carlo
simulation which included geometrical and trigger requirements, was $4250$~m$^2$sr.

No monopole candidates were found. For $ 1.1
\times 10^{-4} < \beta < 5 \times 10^{-3}$ the flux upper limit is $
2.8 \times 10^{-16}$~cm$^{-2}$~s$^{-1}$~sr$^{-1}$ at $ 90 \% $ C.L.
(Fig.~\ref{fig:individuali}, curve ``stream. H").
\begin{figure}
\resizebox{0.46\textwidth}{!}{%
\includegraphics{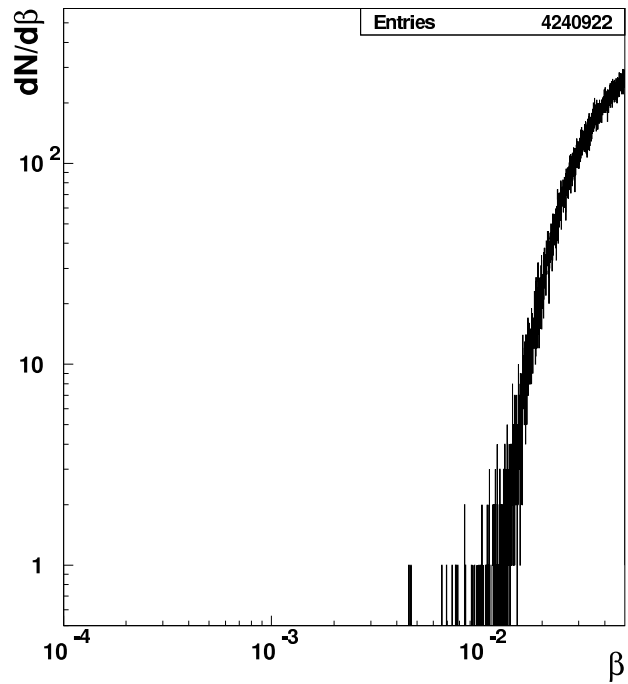}} 
\caption{The $\beta$ distribution for the candidate events reconstructed with the
         vertical streamer tube analysis. Events with $\beta > \beta_{cut}^v=3\times 10^{-3}$
         are muons; all analyzed events are included.}
 \label{streamer1}
\end{figure}

\subsubsection{Search with vertical streamer tubes}
Like in the search with horizontal streamer tubes,  
also in this case, the reconstruction procedure started from time and space information 
provided by the trigger. These were used to select hits compatible in both time
and position with the $\beta$-slice that fired.  
Using these  time-position hits, we performed a complete space and time tracking 
of each events to reject 
accidental alignments; this leaves only the muon background.  This was also used to
evaluate the efficiencies.
The spatial and temporal reconstruction algorithms are the same as those used in the 
horizontal analysis.
On the basis of the overall event reconstruction in the 3 spatial views
($x-z,~d-z,~y-z$) and in the ($t-z$) view, it was possible to achieve the complete
geometrical and temporal reconstruction of the particle track and to compute its $\beta$.
 The $x-z$ spatial view corresponds to the horizontal streamer tube
wire view; the $d-z$ view corresponds to the horizontal streamer tube strips and the 
$y-z$ view corresponds to that of the lateral (vertical) streamer tube wire view.
Finally the $t-z$ view is the temporal view along the vertical direction.
As for the search with horizontal streamer tubes, muons are rejected if 
$\beta >\beta_{cut}^{v}=3\times 10^{-3}$ see Fig.~\ref{streamer1}.

The analysis efficiency was estimated  with the same approach used for the horizontal
streamer monopole search.

The search covered data from October 1994 to September 2000 for a
total of 4.4 liveyears. The  overall average efficiency was $70 \%$. The
acceptance, estimated by Monte Carlo simulation, was $ 3018$~m$^2$sr.
The  measured $\beta$ distribution, shown in Fig.~\ref{streamer1}, is broader
than the one obtained from the horizontal analysis (see \cite{stmono} for comparison). This limited
the sensitivity of this search to the velocity range
$ 1.1 \times 10^{-4} < \beta < 3 \times 10^{-3}$.
No  magnetic monopole candidate was found. 
We thus establish an upper limit to
the monopole flux  $ \Phi \le 8 \times
10^{-16}$~cm$^{-2}$~s$^{-1}$~sr$^{-1}$ for $10^{-4}< \beta <3 \times 10^{-3}$ 
(Fig.~\ref{fig:individuali},
curve ``stream.V").

\subsection{Searches with the Track-Etch detector}
\label{sec:cr39}
The MACRO  nuclear track subdetector \cite{primosm,techpap} covered a total area of 1263~m$^2$ 
and was organized in stacks (``wagons") of $24.5 \times
24.5$ cm$^2$ consisting of three layers of CR39, three layers of
Lexan and an aluminum absorber placed in an aluminized Mylar bag
filled with dry air. 
\begin{figure}
\vspace{0.3cm}
\resizebox{0.45\textwidth}{!}{\includegraphics*{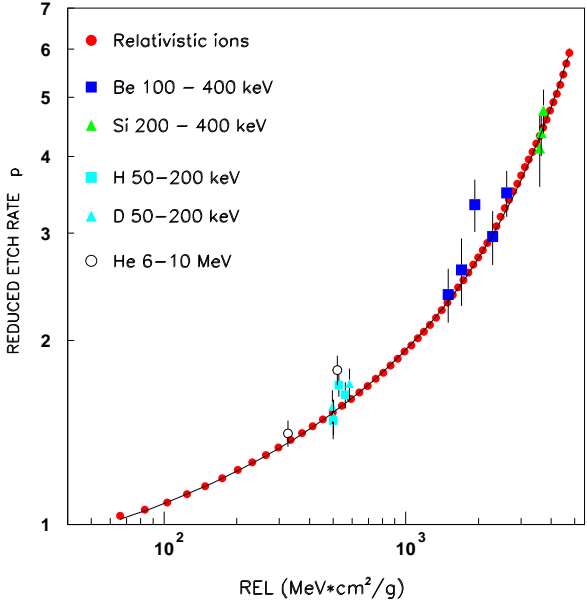}}
\caption{Measured reduced etch rate $p$ vs REL for the CR39 track etch detector exposed
         to relativistic and slow ion beams; the points are the experimental
         data, the solid line is the best fit to the data points ($8 < Z < 68$) obtained with 
         relativistic ions.}
\label{calib}
\end{figure}
The formation of an etchable track in a
nuclear track detector is related to the Restricted Energy Loss
(REL)   \cite{track}. 
There are two contributions to REL: the
electronic energy loss ($S_e$), which represents the energy
transferred to the electrons, and the nuclear energy loss ($S_n$),
which represents the energy transferred to the nuclei in the
material.
In \cite{cr39} it was shown that in  our CR39 
$S_n$ is as
effective as $S_e$ in producing etchable tracks. 
The response of a nuclear  track detector is measured by the
reduced etch rate $p=v_T/v_B$, where $v_T$ and $v_B$ are the
etching rates along the particle track and of the bulk material,
respectively \cite{calib}. The reduced etch rate $p$ vs REL for our 
CR39 is shown in Fig.~\ref{calib}, it was measured using both relativistic
and slow (down to $\beta \sim 4\times 10^{-3}$) ions \cite{cr39}.

The CR39 allowed a search for magnetic monopoles of different 
magnetic charges and $\beta$.  For a single Dirac charge,
our CR39 is sensitive for MM in the ranges $2\times 10^{-5} < \beta <2 \times 10^{-4}$ and 
$\beta >1.2\times 10^{-3}$ \cite{boeloss}.
Lexan has a much higher threshold making it
sensitive to relativistic MMs only ($\beta > 10^{-1}$).

Our CR39 was manufactured by the Intercast
Europe Co. of Parma (Italy). 
A specific production line was set up 
in order to achieve a low detection
threshold, high sensitivity in a large range of energy losses,
high quality of the post-etched surface after prolonged
etching and stability of the detector response over long
periods of time \cite{Produz}.

We analyzed 845.5 m$^2$ of the MACRO CR39
track-etch sub-detector, with an average
exposure time of 9.5 years. 
Since no candidates were found, the Lexan foils were not analyzed.
The top CR39 foils were strongly etched in a 8N NaOH water
solution at 85$^\circ$C till their final thickness reached 
300-400~$\mu m$.
The signal looked for was a hole or a biconical track with
the two base cone areas equal within experimental
uncertainties.

After etching, the foils were scanned twice, using back lighting  by different operators
at low 
magnification looking for any possible optical inhomogeneity;
the double scan guarantees an efficiency close to $100\%$ for finding an  etched hole.
Detected inhomogeneities were further observed with a stereo microscope
and were classified either as surface defects or as particle tracks. 
The latter were further observed under an
optical microscope with high magnification. The axes
of the base-cone ellipses in the top and bottom surfaces of the
foils were measured and the corresponding $p$ and incidence angle
$\Theta$ computed. A track was defined as a candidate if $p$ and
$\Theta$ on the top and bottom sides were equal to within 15\%.
At a residual thickness of 300-400~$\mu m$, double
etch-pit tracks could be  induced by proton recoils from neutron
interactions or by low energy nuclei from muon interactions in the
material surrounding the apparatus.

For the few candidates remaining after the analysis of the
first sheet (an average of 5/m$^{2}$), we looked for a coincidence in position, angles and
RELs in the third (bottom) CR39 layer. 
The second foil was etched in 6N NaOH
water solution at 70 $^{\circ}$C for 30 hours. An accurate
scanning at high magnification was performed
within an area of about 1 mm$^2$ around the expected position of
the candidate. 
 
We estimate a global efficiency  of the procedure of
99$\%$. 
No two-fold coincidence was found, that is no
magnetic monopole  candidate was detected.
The 90\% C.L. flux upper limits are at the level of $2.1
\times 10^ {-16}$ cm$^{-2}$s$^{-1}$sr$^{-1}$ for $g=g_D$ 
magnetic monopoles with
$\beta \sim 10^{-4}$ and $1.5\times 10^ {-16}$
cm$^{-2}$s$^{-1}$sr$^{-1}$ for magnetic monopoles with $\beta \sim 1$
(Fig.~\ref{fig:individuali}, curves ``CR39").

\subsection{Combined search for fast magnetic monopoles}
\label{sec:combin}

A search for fast magnetic monopoles with scintillators or streamer tubes alone was 
affected by
the background due to high energy muons with large energy losses in the detector.
This background could be reduced by a combination of geometrical
and energy cuts imposed on each of the sub-detectors.
A combined analysis, based on all three subsystems, allowed the use of
these cuts in a rather conservative way.
Moreover, any systematic error was greatly reduced by the combination
of measurements from the three subsystems.
The redundancy and complementary offered by the MACRO detector
allowed a good rejection power against the background and
a high reliability of possible candidates.
\begin{figure}
\vspace{-0.3cm}
\resizebox{0.51\textwidth}{!}{ \includegraphics*{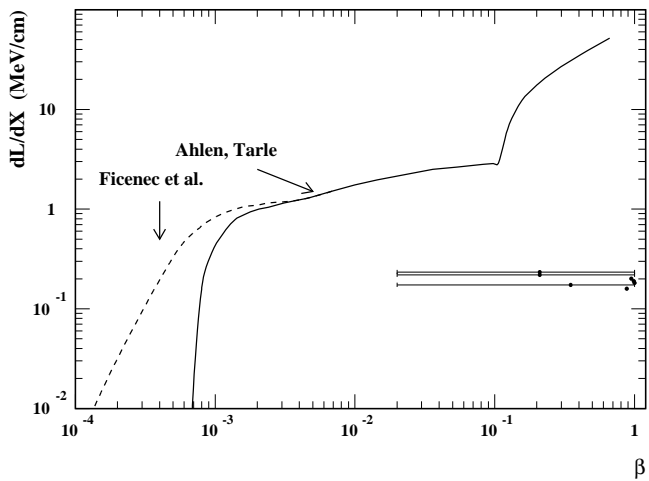}  }
\caption{ Combined search: Light yield for
          the seven  MM candidates
          which survived
          the scintillator and streamer tube cuts. 
          The candidates light yields are compared with the expected magnetic monopole signal 
          as computed in
          \protect\cite{ficenec,ahlentarle} (see Sect.~\ref{sec:combin}).}
\label{fig:comb1}
\end{figure}
The analysis procedure is fully described in \cite{fastcomb}. It used the
data taken with the scintillator and the streamer tube sub-detectors
to identify candidate events.
This was done by reconstructing the energy release (using the scintillators' 
Energy Reconstruction Processor --ERP-- and  the streamer tube Charge and Time Processor
-- QTP \cite{primosm,techpap}) 
and the particle's trajectory
(using the streamer tubes' digital hit information).
Any remaining events 
were then searched for in the track-etch layers as a final tool
for their rejection or confirmation.
The  trigger selection criteria required at least one fired scintillation counter
and 7 hits in the  horizontal streamer planes of the lower subdetector \cite{stmono}.
Once the event tracking was performed, the value of the energy lost in the scintillator
 intercepted by the track
was reconstructed  using the ERP system.
The reconstructed energy in each selected scintillation counter was required to be 
$\Delta E \ge (\Delta E)_{min} = 150$~MeV.
The minimum light yield by a MM in 10 cm pathlength in the
scintillator is $\sim$ 230~MeV. 
A further selection was applied on the streamer tube pulse charge
by using the multiple measurements provided by QTPs along the particle trajectory.

A cut \cite{fastcomb} was applied on the event average streamer
charge, by exploiting the logarithmic dependence of the streamer charge on the primary
ionization \cite{gbation,laser}.

In the analyzed data set, seven events survived the cuts.
For these events the track-etch wagons identified by the streamer tracking
system were extracted and analyzed. No track compatible with the crossing of a
MM was found.
As a further check, for each candidate, 
the measured value of the energy lost in the scintillation counters
was compared with
the expected signal of a  magnetic monopole of the same velocity
\cite{ficenec,ahlentarle}. This comparison is shown in
Fig.~\ref{fig:comb1}. For three of the seven events, the TOF
information was provided by the streamer system alone, since only one 
scintillation counter was present. In this case the error on
the reconstructed velocity  is large because of the limited time resolution  (3~$\mu$s)
of the streamer tubes.
For all the events the measured energy losses were
well below the expectations for  a MM.
\begin{figure}
\vspace{-2.5cm}
\resizebox{0.5\textwidth}{!}{
\includegraphics*{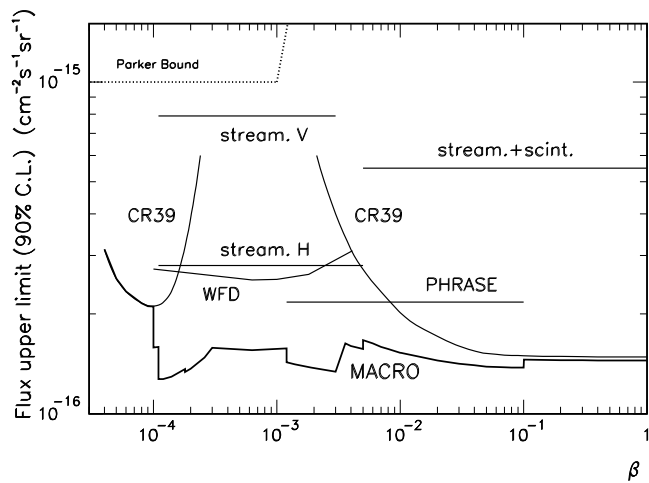}  } 
\caption{Upper limits for an
         isotropic flux of magnetic monopoles obtained by the different MACRO analyses as
         described in the text. The global limit is
         presented as curve ``MACRO".} 
\label{fig:individuali}
\end{figure}
The analysis referred to about 4.8 live years; the maximum geometrical acceptance, 
computed by Monte Carlo
methods, including all the analysis requirements, was 3565 m$^2$ sr. The average global
efficiency was $77\%$.

Since no candidate survived, we set a $90\%$ C.L. flux upper limit at
$5.5 \times 10^{-16}$ cm$^{-2}$ s$^{-1}$ sr$^{-1}$ (presented in Fig.~\ref{fig:individuali} 
as curve ``stream. + scint.") for magnetic monopoles with
$5 \times 10^{-3} < \beta < 0.99$.
The analysis ends at $\beta = 0.99$ since for 
Lorentz factor
$\gamma \gtrsim 10$ magnetic monopoles might induce showers in the detector,
reducing the analysis efficiency \cite{fastcomb}.

\section{The global MACRO limit}
\label{sec:global}
No magnetic monopole candidates were found in any of the above
mentioned searches. Since any one of the sub-detectors may rule out,
within its own acceptance and sensitivity, a potential candidate, a
global limit was computed by combining together the limits obtained
by the single analyses.
The final MACRO limit and the limits obtained by the single
subdetectors are given in Tab.~\ref{tab:limiti} and shown in Fig.~\ref{fig:individuali}.
\begin{table*}
\begin{center}
\begin{tabular}{|c|c|c|c|c|c|c|c|}\hline &
\multicolumn{7}{c|}{Flux upper Limits ($10^{-16}$ cm$^{2}$ s$^{-1}$ sr$^{-1}$)}\\\hline 
$\beta$ range & CR39 & WFD & Stream. H & Stream. V & PHRASE & Stream.+scint. & Global\\
\hline
$(4.0\div 10.0)\times 10^{-5}$ & 3.1$\div$ 2.1  &     &           &           &        &       & 3.1$\div$ 2.1 \\
$(1.0\div  1.1)\times 10^{-4}$ & 2.8          & 2.7 &           &           &        &         & 1.6         \\
$(1.1\div  2.6)\times 10^{-4}$ & 2.2$\div$ 7.5 & 2.5 & 2.8       & 7.9       &        &       & 1.3$\div$ 1.5 \\
$(2.6\div 12.0)\times 10^{-4}$ &              & 2.5 & 2.8       & 7.9       &        &         & 1.6         \\
$(1.2\div  1.9)\times 10^{-3}$ &              & 2.5$\div$2.6 & 2.8  & 7.9       & 2.2   &          & 1.4         \\
$(1.9\div  3.0)\times 10^{-3}$ &7.5$\div$ 3.9  & 2.6$\div$2.9 & 2.8& 7.9       & 2.2   &        & 1.3         \\
$(3.0\div  4.1)\times 10^{-3}$ & 3.9$\div$ 3.1  & 2.9$\div$3.1 & 2.8       & 7.9       & 2.2   &        & 1.6         \\
$(4.1\div  5.0)\times 10^{-3}$ & 3.1$\div$ 2.8  &     & 2.8       &           & 2.2   &  5.5   & 1.6$\div$ 1.66\\
$5.0\times 10^{-3}\div 0.1   $ & 2.8$\div$ 1.5  &     &           &           & 2.2   &  5.5   & 1.8$\div$ 1.5 \\
$0.1\div 1.0                 $ & 1.5          &     &           &           &        &  5.5    & 1.4        \\
\hline
\end{tabular}
\vspace{0.8 cm}
\caption{The 90\% C.L. flux upper limits (in units of $10^{-16}$ cm$^{-2}$ s$^{-1}$ sr$^{-1}$) 
         as a function of $\beta$ for an isotropic flux of $g=g_{D}$ magnetic monopoles with
         $m \ge 10^{17}$ GeV/c$^{2}$. The limits discussed in Section 2 are given in 
         columns two to six; the global MACRO limit discussed in Section 3 is given in
         the last column.}
\label{tab:limiti}
\end{center}
\end{table*}
Each search ``$i$" produced a $90\%$ C.L. flux limit
given by $\Phi_i = 2.3/A_i$, where $A_i$ is the analysis time
integrated acceptance.
In order to obtain the global MACRO limit, we
divided the full $\beta$ interval in a number of slices sufficient
to characterize the changes in the individual acceptances. We
required that the significance of the global limit (in terms of
C.L.) is not altered.

In order to illustrate the algorithm,
suppose that in a specific $\beta$ slice there are
two analyses, ``1" and ``2", based on the use of two sub-detectors
during the same period of time. We consider analysis ``1" as
``dominant" in the sense that it  contributes with its full time
integrated acceptance $A_1$ to the global time integrated
acceptance $A_G$.
Analysis ``2"  then contributes to the global limit only with its
independent parts relative to ``1". We consider both the
temporal independence as well as  the spatial (geometric) independence
versus the dominant analysis.
The temporal independence is determined by comparing the ``time
efficiencies" $\epsilon_i^t$ of the analyses, defined as the
ratios of each analyses live time to the covered solar time. 
If $\epsilon_2^t > \epsilon_1^t$,  the coefficient representing the temporal independence
of ``2" versus ``1" is $c^t_{2,1} = \epsilon_2^t - \epsilon_1^t$;
otherwise, $c^t_{2,1} = 0$.
In the case of the track-etch subdetector, there is no dead-time,
so its temporal efficiency is equal to 1.
The coefficient representing the geometric independence of analysis ``2" versus ``1",
$c^s_{2,1}$, originates from the difference between the
acceptances of the analyses. It is obtained by Monte Carlo
simulations,
assuming an incoming isotropic 
flux of magnetic monopoles with respect to subdetector
``2": $c^s_{2,1} = (N_2 - N_{1,2})/N_2$, where $N_2$ and $N_{1,2}$
are the number of  MMs detected by ``2" and both analyses,
respectively.

The global time integrated acceptance is then:\\
$A_G = A_1 + c^t_{2,1}  A_2 + (1-c^t_{2,1}) c^s_{2,1} A_2$\\
The global $90\%$ C.L. limit for the flux of magnetic monopoles is
$\Phi_G= 2.3/A_G$.

The algorithm used to combine the actual MACRO limits is 
more complicated than the example above. For each analysis we
took into consideration its actual history, eliminating the
longer periods of time in which it was eventually missing, and the
changes in the detector configuration (super-modules
involved). Those corrections were more critical in the case of
earlier analyses, that were carried on during the construction
of the MACRO detector and during initial tests; note that limits
obtained by such older searches are not presented in Fig.~\ref{fig:individuali} 
and in Tab.\ref{tab:limiti}, as
they are considerably higher than the included ones, but they have still their
imprint on the global limit.

\begin{figure}
\resizebox{0.5\textwidth}{!}{
\includegraphics{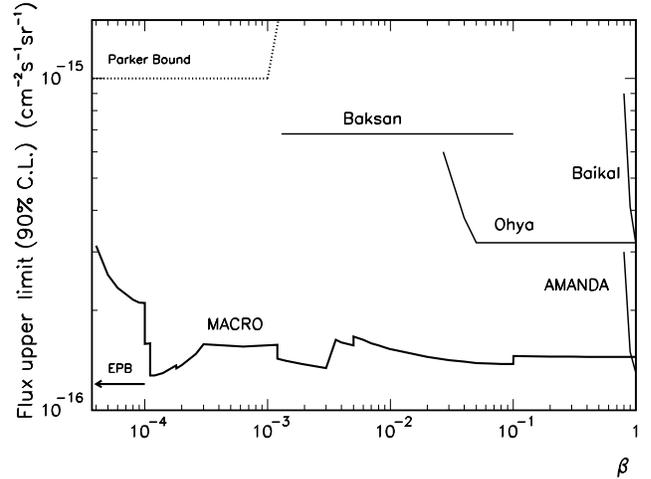}  } 
\caption{The global MACRO
         limit for an isotropic flux of bare magnetic monopoles, with $m
         \geq 10^{17}$ GeV/c$^2$, $g=g_D$ and $\sigma_{cat} <$ few mb. For
         comparison, we present also the flux limits from other
         experiments \cite{otherexp}.} 
\label{fig:global}
\end{figure}

In Fig.~\ref{fig:global} we present the global MACRO limit; for
comparison,  the flux limits from other experiments which searched for
magnetic monopoles with similar properties, are also shown
\cite{otherexp}.
In the figure the arrow indicates the Extended Parker Bound (EPB) at the level of 
$1.2\times10^{-16}$ $(m/10^{17})~$cm$^{-2}$s$^{-1}$sr$^{-1}$, which was obtained by 
considering the survival probability of a magnetic monopole of mass $m$ 
in an early magnetic seed field \cite{epb}.

\section{Discussion}
\label{sec:discuss}

Our analysis applies to an isotropic
flux of bare MMs with charge $g=g_D$=e/2$\alpha$ 
and nucleon decay catalysis cross sections
smaller than 1$\,$mb \cite{monorev}.

The magnetic monopole flux at the detector site is isotropic if magnetic monopoles have enough
kinetic energy (i.e. large mass and/or  high $\beta$) in order to cross the Earth.
If this is not the case, only a fraction of the total detector acceptance is actually
exploited and the upper limits given above must be corrected accordingly.
In particular, if monopoles have sufficient energy to cross the  overburden mountain 
(i.e. $\sim 3700 \,$hg/cm$^2$) and reach 
the detector from above, 
but not enough to cross the Earth,
the flux upper limit  is about a factor of two weaker.
Therefore the energy losses suffered by magnetic monopoles in traversing 
the Earth 
set the accessible values of mass and $\beta$ for a given experiment.
For MACRO, at least one half of the geometrical acceptance is ensured
for relativistic  MMs with $m \gtrsim 10^{6}\,$GeV/c$^{2}$ or, for 
$\beta \simeq 10^{-4}$ if $m \gtrsim 10^{10}\,$GeV/c$^{2}$; 
the full acceptance is reached if
$m \gtrsim 10^{10}\,$GeV/c$^{2}$ or $m \gtrsim 10^{16}\,$GeV/c$^{2}$,
for relativistic and slow magnetic monopoles, respectively \cite{idmthesis,boearth} 
(see Table 2).
Sensitivity to fast  and light MMs is also important, since they could be 
responsible for ultra high energy cosmic ray events above the
Greisen-Zatsepin-Kuzmin cutoff \cite{supergzk}.

If magnetic monopoles have $g > g_D$ and/or an associated electric charge
(dyons \cite{monorev}), the detection efficiency might change.
As far as our sub-detectors are concerned,
the dyon detection efficiency would be greater than that
of bare monopoles since the excitation/ ionization-based energy losses would be
larger \cite{boeloss}.
The only exception is the detection of very slow dyons with the streamer tubes,
since it relies on the Drell effect on helium atoms \cite{drellpate}, which for dyons
might be prevented by coulombian repulsion.
However, as shown in \cite{boeloss}, the only effect would be 
the raising of the minimum
velocity threshold at $\beta \simeq (1.7 \times 10^{-4})\sqrt{Z}$ for dyons with
electric charge $Ze$, the threshold for bare  MMs being
$\beta = 1.1 \times 10^{-4}$.
\begin{table}
\begin{center}
\begin{tabular}{|c|c|c|}\hline
$\Phi$ (cm$^{2}$ s$^{-1}$ sr$^{-1}$) & m (GeV/c$^{2}$)   & $\beta$ \\ \hline
$\le 1.4\times 10^{-16}$            & $\gtrsim 10^{10}$ & $>10^{-1}$\\
$\le 1.4\times 10^{-16}$            & $\gtrsim 10^{16}$ & $>10^{-4}$\\
$\le 2.8\times 10^{-16}$             & $\gtrsim 10^{6}$  & $>10^{-1}$\\
$\le 2.8\times 10^{-16}$             & $\gtrsim 10^{10}$ & $>10^{-4}$\\\hline
\end{tabular}
\vspace{0.5 cm}
\caption{Final results of magnetic monopole searches with the MACRO experiment.
         The limits depend on the magnetic monopole mass and on its velocity.}
\label{tab:finale}
\end{center}
\end{table}
As suggested in \cite{rubacallan}, GUT magnetic monopoles may
catalyze nucleon decays along their path with a cross section
$\sigma_c$ of the order of the hadronic cross sections
\cite{monorev}. If $\sigma_c \gtrsim 1\,$mb, the efficiencies of
the aforementioned searches might decrease due to the effects of
the fast decay products. 
A deep study was performed on the
effect of the nucleon catalysis cross section on the streamer tube analysis.
The main result is that the present analyses are still efficient up to at least
$\sigma_c \simeq 100\,$mb. A dedicated  search  
for  magnetic monopoles accompanied
by one or more nucleon decays along their path  was also performed. 
The results are
reported in a separate paper \cite{catalysis}.

\section{Conclusions}
\label{sec:conclu}

We present the final results of GUT magnetic monopole searches performed with the MACRO detector
at Laboratori Nazionali del Gran Sasso (Italy). 
Different searches using the MACRO sub-detectors (i.e. scintillation counters,
limited streamer tubes and nuclear track detectors) both 
in stand alone and combined ways, were performed.
Since no candidates were detected, Tab.2 summarizes the $90\%$ C.L upper limit to an isotropic
flux of bare MMs with charge $g=g_D$=e/2$\alpha$ 
and nucleon decay catalysis cross section
smaller than 1 mb.

\vskip 1.cm
{\bf Acknowledgements} \\
We gratefully acknowledge the support of the director and of the staff of the Laboratori
Nazionali del Gran Sasso and the invaluable assistance of the technical staff of the
Institutions participating in the experiment. We thank the Istituto Nazionale di Fisica
Nucleare (INFN), the U.S. Department of Energy and the U.S. National Science Foundation
for their generous support of the MACRO experiment. We thank INFN, ICTP (Trieste),
WorldLab and NATO for providing fellowships and grants for non Italian citizens.

\vskip 1.cm



\begin{thebibliography}{}

\bibitem{monorev}    J. Preskill, Ann. Rev. Sci. \textbf{34} (1984) 461;\\ 
                     G. Giacomelli, Riv. Nuovo Cimento \textbf{7} (1984) 1;\\ 
                     D. E. Groom, Phys. Rep. \textbf{140} (1986) 324 
                     and references therein.
\bibitem{shafi}      S. F. King and Q. Shafi, Phys. Lett. \textbf{B422} (1998) 135;\\ 
                     T. W. Kephart and Q. Shafi, hep-ex/0105237 (2001). 
\bibitem{lightMM}    T. W. Kephart and T. J. Weiler, Astrop. Phys. \textbf{4} (1996) 217;\\
                     C. O. Escobar and R. A. Vazquez, Astrop. Phys. \textbf{10} (1999) 197. 
\bibitem{parker}     M. S. Turner et al., Phys. Rev. \textbf{D26}
                     (1982) 1296.
\bibitem{primosm}    S. P. Ahlen et al. (MACRO Coll.),
                     Nucl. Instr. $\&$ Meth. in Phys. Res. \textbf {A324} (1993) 337.
\bibitem{techpap}    M. Ambrosio et al. (MACRO Coll.), NIM {\bf A486} (2002) 663.
\bibitem{trd}        M. Ambrosio et al. (MACRO Coll.),
                     Astropart. Phys. \textbf{10} (1999) 11, hep-ex/9807009.
\bibitem{boeloss}    J. Derkaoui et al., Astropart. Phys. \textbf {10} (1999) 339.
\bibitem{monopap1}   M. Ambrosio et al. (MACRO Coll.), Phys. Lett.
                     \textbf{B406} (1997) 249.
\bibitem{hong}       S. P. Ahlen et al (MACRO Coll.), Phys. Rev. Lett.
                     \textbf{72} (1994) 608.
\bibitem{phraserp}   M. Ambrosio et al.(MACRO coll.), Astropart. Phys.
                     \textbf{6} (1997) 113.
\bibitem{stmono}     M. Ambrosio et al. (MACRO coll.), Astropart. Phys.
                     \textbf{4} (1995) 33.
\bibitem{track}      L. Patrizii (MACRO Coll.), Proc. of the XXth ICNTS,
                     Rad. Meas. \textbf{34} (2001) 259.
\bibitem{fastcomb}   M. Ambrosio et al. (MACRO Coll.), hep-ex/0110083,
                     accepted by Astroparticle Physics.
\bibitem{ficenec}    D. J. Ficenec et al., Phys. Rev. \textbf{D36} (1987) 311.
\bibitem{Ahlen1}     S. P. Ahlen, Phys. Rev. \textbf{D17} (1978) 229.
\bibitem{Ahlen2}     S. P. Ahlen and K. Kinoshita, Phys. Rev. \textbf{D26} (1982) 2347.
\bibitem{ahlentarle} S. P. Ahlen and G. Tarl\'e, Phys. Rev. \textbf{D27} (1983) 688.
\bibitem{sophia}     S. Kyriazopoulou (MACRO Coll.), Proc. NATO ARW 
                     on Cosmic Radiations, Oujda (Morocco) 21-23 March 2001;\\
                     S. Kyriazopoulou, Ph.D. Thesis, California Institute of Tecnology (2002).
\bibitem{collasso}   S. P. Ahlen et al. (MACRO Coll.), Astropart. Phys. 
                     \textbf{1} (1992) 11.
\bibitem{sorgente}   A. Baldini et al., Nucl. Instr. \& Meth. in Phys. Res.
                     \textbf{A305} (1991) 475.
\bibitem{drellpate}  S. D. Drell et al., Phys. Rev. Lett. \textbf{50} (1983) 644. 
                     V. Patera, Phys. Lett. \textbf{A137} (1989) 259.
\bibitem{gbation}    G. Battistoni et al.,
                     Nucl. Instr. $\&$ Meth. in Phys. Res. \textbf{A270} (1988) 185.
\bibitem{laser}      G. Battistoni et al.,
                     Nucl. Instr $\&$ Meth. in Phys. Res. \textbf{A401} (1997) 309.
\bibitem{ivan}       I. De Mitri et al., Nucl. Instr. \& Meth. in Phys. Res. 
                     \textbf{A360} (1995) 311.
\bibitem{muon}       G. Battistoni et al.,
                     Nucl. Instr $\&$ Meth. in Phys. Res. \textbf{A399} (1997) 244.
\bibitem{cr39}       S. Cecchini et al.,  Riv. Nuovo Cimento
                     \textbf{A109} (1996) 1119.
\bibitem{Produz}     S. Cecchini et al. Radiat. Meas., \textbf{34} (2001) 54.
\bibitem{calib}      G. Giacomelli et al.,  Nucl. Instr $\&$ Meth. in Phys. Res. 
                     \textbf{A411} (1998) 41.
\bibitem{icrc2001}   B. Choudhary (MACRO Coll.), 26$^{th}$ ICRC, Salt Lake City, Utah, USA (1999),
                     hep-ex/9905023;\\ 
                     M. Sitta (MACRO Coll.), 27$^{th}$ ICRC, Hamburg, Germany (2001).
\bibitem{otherexp}   E. N. Alexeyev et al. (``Baksan''), 21$^{st}$ ICRC, Adelaide, Vol. 10, (1990) 83;\\
                     S. Orito et al. (``Ohya''), Phys. Rev. Lett. \textbf{66} (1991) 1951;\\
                     V. A. Balkanov for the Baikal Coll. (``Baikal'', ``AMANDA''), Nucl. Phys. 
                     \textbf{B91} (2001) 438;\\
                     G. Giacomelli et al., hep-ex/0005041 (2000).
\bibitem{epb}        F.C. Adams et al., Phys. Rev. Lett. \textbf{70} (1993) 2511.
\bibitem{idmthesis}  I. De Mitri, Ph.D. Thesis, University of L'Aquila (1996).
\bibitem{boearth}    J. Derkaoui et al., Astropart. Phys. \textbf {9} (1998) 173.
\bibitem{supergzk}   T. W. Kephart and T. J. Weiler, Astropart. Phys. \textbf{4} (1996) 271;\\ 
                     P. Bhattacharjee and G. Sigl, Phys. Rep. \textbf{327} (2000) 109;\\ 
                     S. D. Wick et al., astro-ph/0001233 
                     and references therein.
\bibitem{rubacallan} V. A. Rubakov, JETP Lett. \textbf{33}(1981) 644;\\ 
                     C. G. Callan, Phys. Rev. \textbf{D25} (1982) 2141.
\bibitem{catalysis}  M. Ambrosio et al. (MACRO Coll.), {\it Search for Nucleon Decay induced by GUT Magnetic Monopoles with the MACRO Experiment}, hep-ex 0207024.

\end{thebibliography}
\end{document}